\documentclass[twocolumn,epjc3]{svjour3} 
\RequirePackage[T1]{fontenc}
\smartqed
\RequirePackage{graphicx}
\RequirePackage{mathptmx}
\RequirePackage{flushend}
\RequirePackage[numbers,sort&compress]{natbib}
\RequirePackage[colorlinks,citecolor=blue,urlcolor=blue,linkcolor=blue]{hyperref}
\usepackage{amsmath}
\usepackage{subfig}
\usepackage{float}
\usepackage{balance}
\journalname{Eur. Phys. J. C}
\begin{document}

\title{Hawking emission of charged particles from an electrically charged spherical black hole with scalar hair}

\author{Avijit Chowdhury\thanksref{e1,addr1}}
 \thankstext{e1}{e-mail: ac13ip001@iiserkol.ac.in}
\institute{IISER Kolkata, Mohanpur Campus, Mohanpur, Nadia, 741246, India \label{addr1}}
\date{}
\maketitle
\begin{abstract}
A static spherically symmetric black hole usually turns out to be either a Schwarzschild black hole or a Reissner-Nordstr\"{o}m black hole. This result was summarised by Ruffini and Wheeler as the so-called no hair conjecture which states that for a spherically symmetric black hole only the information about mass ($M$) and electric charge ($e$) of the black hole is available for an external observer.  In this work, we calculate the emission rate of charged particles from an asymptotically flat charged spherically symmetric black hole endowed with a scalar hair using a semi-classical tunneling formalism. We observe that the total entropy of the black hole contains an energy-dependent part due to the scalar charge. The upper bound on the charge-mass ratio of the emitted particles is also observed to decrease with the scalar charge as well.

\end{abstract}

\maketitle
\section{Introduction}
Black holes are fascinating offshoots of general relativity, unmatched in their elegance and complexity. The recent detections of gravitational waves by LIGO and Virgo\cite{LIGO_GW150914, LIGO_GW170104, LIGO_GW151266, abbott_APJ_2017, LIGO_GW170814, LIGO_P1800324, LIGO_P1800307} and the recently obtained ``image" of M87* black hole at the centre of the M87 galaxy  by the Event Horizon Telescope\cite{EHT_APJL_2019} have quite conclusively established that black holes are not merely mathematical constructs but actual physical entities. 
 In 1974, Hawking\cite{hawking_CMP_1975} showed that ``black holes'' are not ``entirely black''; they spontaneously emit particles at a temperature proportional to their surface gravity. Since then various derivations of Hawking radiation have been proposed. In particular, Parikh and Wilczek\cite{parikh_PRL_2000} used the semi-classical tunneling formalism to study Hawking emission of massless uncharged particles  from the Schwarzschild and Reissner-Nordstr\"{o}m (RN)  black holes. According to this formalism, the black hole loses energy due to radiation and thereby decreases its horizon radius. Thus, the outgoing particle creates a potential barrier by itself (see Ref.~\cite{parikh_IJMPD_2004, parikh_GRG_2004,  parikh_MARCELGROSSMANN10}). Using WKB approximation one can evaluate the emission rate from the imaginary part of the action of the outgoing particles. The tunneling method has since been used to study Hawking radiation from a wide variety of black holes. Zhang and Zhao\cite{zhang_JHEP_2005} extended the  tunneling method to study the emission of massive charged particles from Reissner-Nordstr\"{o}m black holes. Later, Jiang and Wu\cite{jiang_PLB_2006} used tunneling formalism to study Hawking radiation of charged particles from Reissner-Nordstr\"{o}m-de Sitter black hole. Tunneling of charged particles in the modified RN black hole was studied by Liu \cite{liu_IJTP_2014}. Jiang,Wu and Cai\cite{jiang_PRD_2006} studied the tunneling of particles from Kerr and Kerr-Newman black holes. Jiang, Yang and Wu\cite{jiang_IJTP_2006} also studied the tunneling of charged particles from Reissner-Nordstr\"{o}m black holes of arbitrary  dimensions. Jiang and Wu\cite{jiang_PLB_2008} studied the tunneling of charged particles from Reissner-Nordstr\"{o}m-de Sitter black holes with a global monopole. Sarkar and Kothawala\cite{sarkar_PLB_2008} gave a generalised treatment of Hawking radiation as tunneling for asymptotically flat, spherically symmetric black holes. Tunneling mechanism has also been extensively used to study Hawking radiation from numerous other black holes (for example, see Refs.~\cite{wu_JHEP_2006, li_PLB_2006, chen_PLB_2008, chen_CQG_2008}). Tunneling of Dirac particles from black rings was studied in Ref.~\cite{jiang_PRD_2008}. Jiang, Chen and Wu\cite{jiang_JCAP_2013} also studied the tunneling of massive particles from the cosmological horizon of a Schwarzschild-de Sitter black hole.The general outcome of all these investigations is the non-thermality of the Hawking emission spectrum. For a detailed review of the tunneling mechanism, we refer the reader to Ref.~\cite{vanzo_CQG_2011}.

The question of information loss has also been addressed using the tunneling formalism. Using logarithmic correction to the Bekenstein-Hawking entropy, Chen and Shao\cite{chen_PLB_2009} showed that in a black hole evaporation process unitarity is preserved and the information loss paradox can be successfully resolved. Singleton \textit{et al.}\cite{singleton_JHEP_2010} showed that if the back reaction and the quantum corrections are taken into account, the information can be carried away by the correlations of the outgoing radiation during complete evaporation of the black hole and the information loss paradox can be solved. Sakalli \textit{et al.}\cite{sakalli_ASS_2012}  studied hawking radiation in  linear dilaton black holes and showed that no  information is lost during complete evaporation of the black hole. 

In the present work, we employ the tunneling method to study Hawking radiation of charged particles from an asymptotically flat static spherically symmetric charged black hole, endowed with a scalar hair, dubbed as the scalar-hairy RN black hole\cite{astorino_PRD_2013}. The scalar charge comes as an additive correction to the square of electric charge and can survive even in the absence of the electric field. For negative values of the scalar charge in the absence of the electric charge, or even if the negative scalar charge exceeds in magnitude the square of the electric charge, the metric reduces to that of a mutated Reissner-Nordstr\"{o}m spacetime, leading to an Einstein-Rosen bridge\cite{rosen_PR_1935} . The quasinormal modes of scalar and Dirac fields for this spacetime have been studied by Chowdhury and Banerjee\cite{paper1_EPJC_2018} for both positive and negative values of the scalar charge. The superradiant stability of this spacetime has been studied in Ref.~\cite{paper2_arxiv_2019}. The primary objective of the present work is to  study the effect of the scalar hair on the transmission rate of charged Hawking quanta. Based on the non-negativity of mutual information between consecutive Hawking emissions (see Ref.~\cite{zhang_PLB_2009, kim_PLB_2014}), we also find the dependence of the charge to mass ratio of the emitted particles on the scalar charge $s$ in each step.

We observe that the change in entropy of the scalar-hairy RN black hole due to the emission of charged particles contains an additional energy-dependent contribution which vanishes during emission of uncharged particles and also in the absence of the scalar hair. The tunneling rate  for the emission of charged particles from the scalar-hairy RN black hole matches smoothly with that of the standard RN black hole, in the absence of the scalar hair.

The paper is organised  as follows. We start with a brief description of the background spacetime in Sec.\ref{sec:scalar-hairy RN}. Sec.\ref{sec:tunneling rate} deals with the derivation of the emission rate of charged scalar particles from the scalar-hairy RN  black hole. In Sec.\ref{sec:charge-mass_ratio}, we show the dependence of the upper bound of the charge-mass ratio of the particles, emitted in each step, on the scalar charge.  Finally, in Sec.\ref{sec:summary} we discuss the results that we arrive at.  

\section{The Scalar-hairy-Reissner-Nordstr\"{o}m black hole}\label{sec:scalar-hairy RN}
We consider an  action in which gravity is coupled to Maxwell field $F^{\mu \nu}$ and conformally coupled to a scalar field $\psi$,
\begin{equation}\label{eq_action}
\begin{split}
I=\frac{1}{16\pi G}\int d^4 x\sqrt{-g}\left[ R-F_{\mu \nu}F^{\mu \nu}\right.\\ \left.-8\pi G\left(\bigtriangledown_\mu \psi \bigtriangledown^\mu \psi+\frac{R}{6}\psi^2\right)\right].
\end{split}
\end{equation}
The Einstein equations obtained by extremizing this action admits  an asymptotically flat  Reissner-Nordstr\"{o}m type solution, endowed with a scalar hair $s$~\cite{astorino_PRD_2013},
\begin{equation}\label{eq_metric}
ds^2=-f\left(r \right)dt_R^2+{f\left(r \right)}^{-1}dr^2+r^2 \left( d\theta^2+\sin^2{\theta} d\phi^2 \right),
\end{equation}
with 
\begin{eqnarray}
\label{eq_f(r)}
f\left( r \right)&=&\left(1-\frac{2M}{r}+\frac{e^2+s}{r^2}\right),\\
\psi&=&\pm \sqrt{\frac{6}{8 \pi G}}\sqrt{\frac{s}{s+e^2}}~,
\end{eqnarray}
where $M$ and $e$ are respectively the mass parameter and electric charge of the black hole.
The total energy-momentum tensor of the electromagnetic field and the scalar field is given by
\begin{equation}\label{eq_stressenergy}
T^\mu_\nu=\frac{e^2+s}{r^4} diag\left(-1,-1,1,1\right). 
\end{equation}
The scalar field $\psi$ comes out to be a constant; however, its contribution to the geometry is non-trivial. It is interesting to note that the scalar field $\psi$ can survive independently of the electromagnetic field; hence, the scalar hair is a primary hair. Further, the energy-momentum tensor due to the scalar field being traceless, the existence of this hair is consistent with the  theorem given in Ref.~\cite{narayan_pramana_2015}. 

The energy-momentum tensor of the scalar-hairy RN spacetime satisfies both the dominant and strong energy conditions for $s>-e^2.$ For this work, we will assume $s>0$, so this condition is satisfied. Henceforth, we will also assume $G=c=k_B=\hbar=1.$

The scalar-hairy RN spacetime, Eq.\ \eqref{eq_metric}, is characterised by an event horizon at $r_+$ and a Cauchy horizon at $r_{-}$ where
\begin{equation}\label{eq_r+-}
r_\pm=M \pm \sqrt{M^2-e^2-s}~.
\end{equation}
The spacetime suffers from coordinate singularities both at the event horizon at $r_+$ and at the inner horizon at $r_-$~. To eliminate these singularities at $r_{\pm}$, we introduce the generalised Painlev\'{e} transformation,
 \begin{equation}\label{eq_painleve_trans}
 \begin{split}
 t&= t_R+2\sqrt{2Mr-e^2-s}+M \ln\left( \frac{r-\sqrt{2Mr-e^2-s}}{r+\sqrt{2Mr-e^2-s}} \right) \\
& +\frac{e^2+s-2M^2}{\sqrt{M^2-e^2-s}} \tanh^{-1}\left(\frac{\sqrt{M^2-e^2-s}\sqrt{2Mr-e^2-s}}{Mr-e^2-s}\right),
\end{split}
\end{equation}  
which transforms the scalar-hairy RN metric (Eq.\ \eqref{eq_metric}) to
\begin{equation}\label{eq_painleve_metric}
\begin{split}
ds^2= &-f(r)dt^2+2\sqrt{1-f(r)} dt dr+dr^2+r^2 d\theta^2\\
&+r^2\sin^2{\theta} d\phi^2 .
\end{split}
\end{equation}
The line element in Eq.\ \eqref{eq_painleve_metric} highlights the static but non-stationary character of the spacetime. We will use this line element to study the emission of massive charged particles from the scalar-hairy RN black hole  using the semi-classical tunneling formalism (see Ref.~\cite{parikh_PRL_2000, parikh_IJMPD_2004, zhang_JHEP_2005}).

\section{Tunneling rate of charged particles}\label{sec:tunneling rate}
The tunneling of particles across a potential barrier being an instantaneous phenomenon, the metric must obey Landau's condition of coordinate clock synchronisation\cite{landau_vol2}. During tunneling two events occur simultaneously, a radially moving particle tunnels into the barrier while another particle tunnels out of the barrier. According to Landau's theory of coordinate clock synchronisation, the coordinate time difference between two simultaneous events, occurring at two different space points is given by \cite{zhang_JHEP_2005},
\begin{equation}
dt=-\frac{g_{01}}{g_{00}}dr_c~, \quad \left( d \theta=d \phi =0 \right),
\end{equation}
where $r_c$ is the position of the tunneling particle.
Following \cite{zhang_JHEP_2005, zhang_PLB_2005, jiang_PRD_2006,  jiang_PLB_2006, ali_IJTP_2014, ali_IJTP_2008}, we consider the outgoing charged particle to be represented by a de Broglie s-wave whose phase velocity $v_p$ is related to its group velocity $v_g$ as,
\begin{equation}\label{eq_vp_vg}
v_p=\frac{v_g}{2}~.
\end{equation}
The group velocity of a de Broglie s-wave representing an outgoing charged Hawking quanta is given by,
\begin{equation}\label{eq_groupvel}
v_g=\frac{dr_c}{dt}=-\frac{g_{00}}{g_{01}}=\frac{e^2-2 M r+r^2+s}{r \sqrt{-e^2+2 M r-s}}~,
\end{equation}
which results in a phase velocity of
\begin{equation}\label{eq_vp}
\dot{r}=v_p=\frac{e^2-2 M r+r^2+s}{2r \sqrt{-e^2+2 M r-s}} ~.
\end{equation}
The electric charge of the black hole gives rise to an electromagnetic field  $F_{\mu\nu}$, given by the vector potential $A_\mu=-\delta^0_\mu \left(e/r \right)$. The Lagrangian of this electromagnetic field is $L_e=-\frac{1}{4} F_{\mu\nu} F^{\mu\nu}$. However, $L_e$ being independent of the corresponding generalised coordinates $A_\mu=\left( A_t,0,0,0 \right)$, there exists a gauge freedom in the choice of $A_t$. To eliminate this freedom, we write the action of an outgoing charged massive particle as\cite{landau_vol2, zhang_JHEP_2005} 
\begin{equation}
\mathcal{A} = \int^{t_f}_{t_i} \left(L-P_{A_t}\dot{A_t}\right) dt~,
\end{equation}
where $P_{A_t}$ is the canonical momentum conjugate to the generalised coordinate $A_t$ and $L$ is the total Lagrangian of the matter-gravity system.
When a particle of charge $q$ and mass $\omega$ tunnels out of the event horizon the electric charge of the black hole changes from $e\rightarrow e-q$ and the ADM mass changes from $M_{ADM}\rightarrow M_{ADM}-\omega$. The imaginary part of the corresponding action can thus be written as
\begin{equation}\label{eq_Ims}
\begin{split}
Im \hspace{2pt} \mathcal{A} & = Im \left\lbrace \int^{r_f}_{r_i} \left( P_r \dot{r}-P_{A_t}\dot{A_t}\right)\frac{dr}{\dot{r}} \right\rbrace \\
& = Im \left\lbrace \int^{r_f}_{r_i} \left[ \int^{\left( P_r P_{A_t} \right)}_{\left(0,0\right)} \left( d P'_r \dot{r}-\dot{A_t} dP'_{A_t} \right) \right] \frac{dr}{\dot{r}} \right\rbrace,
\end{split}
\end{equation}
where $r_i$ and $r_f$ are the position of the event horizon before and after the tunneling of the charged Hawking quanta. Using Hamilton's equations,
\begin{eqnarray}
\dot{r} &=& \frac{d H}{d P_r}\bigg\vert_{\left( r; A_t,P_{A_{t}} \right)}\\
\mbox{and} \quad \dot{A_t} &=& \frac{d H}{d P_{A_{t}}}\bigg\vert_{\left( A_{t}; r,P_r \right)},
\end{eqnarray}
with
\begin{eqnarray}
& dH\vert_{\left( r; A_t,P_{A_{t}} \right)} = d \left(M_{ADM}-\omega'\right)=-d\omega' \\
\mbox{and} \quad & dH\vert_{\left( A_{t}; r,P_r \right)} = \frac{e-q'}{r}d(e-q')=-\frac{e-q'}{r} dq',
\end{eqnarray}
we rewrite the Eq.\ \eqref{eq_Ims} as
\begin{equation}\label{eq_Ims1}
Im \hspace{2pt} \mathcal{A}=-Im \left[ \int^{r_f}_{r_i}  \left( \int^{\omega}_0  d \omega' - \int^q_0 \frac{e-q'}{r} d q'\right)\frac{dr}{\dot{r}} \right].
\end{equation}

The ADM mass $M_{ADM}$ is related to the  mass parameter $M$ as
\begin{equation}\label{eq_MADM}
M_{ADM}=\frac{M}{1+s/e^2}~.
\end{equation} 
Since $s>-e^2$,  $M_{ADM}$ is always positive definite and for the standard RN black hole $(s=0)$, it is equal to the mass parameter $M.$ However, in the limit of $e\rightarrow	0$, $M_{ADM}$ vanishes. 

Substituting Eq.\ \eqref{eq_MADM} in the expression of $\dot{r}$ in Eq.\ \eqref{eq_vp}  with $M_{ADM}\rightarrow M_{ADM}-\omega'$ and $e\rightarrow e-q'$ we get
\begin{equation}\label{eq_vp_final}
\dot{r}=\frac{-2 r (M_{ADM}-\omega' ) \left(\frac{s}{(e-q')^2}+1\right)+(e-q')^2+r^2+s}{2 r \sqrt{2 r (M_{ADM}-\omega' ) \left(\frac{s}{(e-q')^2}+1\right)-(e-q')^2-s}}~.
\end{equation}
Eq.\ \eqref{eq_Ims1} in conjunction with Eq.\ \eqref{eq_vp_final} yields
\begin{equation}\label{eq_ImA}
\begin{split}
Im \hspace{2pt} \mathcal{A} &=-Im \Bigg[ \int^{r_f}_{r_i} \left( \int^\omega_0 d \omega' - \int^q_0 \frac{e-q'}{r} d q'\right) \\
& \frac{2 r \sqrt{2 r (M_{ADM}-\omega' ) \left(\frac{s}{(e-q')^2}+1\right)-(e-q')^2-s}}{\left(r-r'_+\right) \left(r-r'_-\right)} dr \Bigg]~,
\end{split}
\end{equation}
where
\begin{eqnarray}
\begin{split}
&r'_{\pm} = (M_{ADM}-\omega' ) \left(\frac{s}{(e-q')^2}+1\right)\\ 
 & \pm \sqrt{(M_{ADM}-\omega' )^2 \left(\frac{s}{(e-q')^2}+1\right)^2-(e-q')^2-s}~,
\end{split}\\
\label{eq_ri}
r_i = M_{ADM} \left(\frac{s}{e^2}+1\right)+\sqrt{M_{ADM}^2 \left(\frac{s}{e^2}+1\right)^2-e^2-s}~, \\
\begin{split}
&r_f = (M_{ADM}-\omega ) \left(\frac{s}{(e-q)^2}+1\right)\\ 
 &+\sqrt{(M_{ADM}-\omega )^2 \left(\frac{s}{(e-q)^2}+1\right)^2-(e-q)^2-s}~.
\end{split}
\end{eqnarray}
Interchanging the order of integrations in \eqref{eq_ImA}, we note that the $r$ - integral has a pole at $r'_+$. Deforming the contour around this pole we obtain,
\begin{equation}\label{integral_inexact}
Im\hspace{2pt} \mathcal{A} =\pi \left( \int^{\omega}_0 \frac{2 {r'_+}^2}{r'_+-r'_-} d\omega'- \int^q_0 \frac{2 r'_+\left(e-q'\right)}{r'_+-r'_-} dq'\right).
\end{equation} 

The First law of black hole mechanics for a scalar-hairy RN black hole is
\begin{equation}\label{eq_1st law}
d M_{ADM}=\frac{\kappa}{8 \pi \left(1+s/e^2 \right)}d A +\Phi \left[ 1+\frac{s}{{r_{-}}^2}\right] de ~,
\end{equation}
where $A$ is the area of the event horizon, $\kappa$ is the surface gravity and $\Phi$ is the electric potential at the horizon,
\begin{equation}
A= 4 \pi {r_+}^2~, \quad \kappa=\frac{r_{+}-r_{-}}{2 {r_{+}}^2}~, \quad \Phi=\frac{e}{r_{+}}~ .
\end{equation}

Using the First law \eqref{eq_1st law}, the integral in \eqref{integral_inexact} can be written as,
\begin{equation}
Im\hspace{2pt} \mathcal{A} =-\frac{\pi}{2} \int^{\left(\omega,q\right)}_{\left(0,0\right)}d\left(\frac{{r'_{+}}^2}{1+\frac{s}{\left(e-q' \right)^2}}\right) -\frac{\Delta S_{\omega}}{2}~,
\end{equation}
where
\begin{equation}\label{eq_Somega}
\Delta S_{\omega} =-2 \pi s \int^q_0 \frac{\left( e-q' \right)}{r'_-{^2}} \frac{r'_+ + r'_-}{r'_+-r'_-}dq'.
\end{equation}
Thus, we get 
\begin{eqnarray}\label{eq_ImS_corr}
Im\hspace{2pt} \mathcal{A} &=&  -\frac{\pi}{2} \left[ \frac{r_f^2}{1+\frac{s}{\left(e-q\right)^2}} - \frac{r_i^2}{1+\frac{s}{e^2}} \right]-\frac{\Delta S_{\omega}}{2}\\
&=&-\frac{1}{2}\left( \Delta S_{BH}+\Delta S_{\omega}\right),
\end{eqnarray}
where 
\begin{equation}\label{eq_Sbh}
\Delta S_{BH}=\left( \pi\frac{r_f^2}{1+\frac{s}{\left(e-q\right)^2}} -\pi \frac{r_i^2}{1+\frac{s}{e^2}} \right)~,
\end{equation}
is the change in the Bekenstein-Hawking entropy\cite{hawking_PRL_1971, bekenstein_PRD_1973} of the scalar-hairy RN black hole, $S_{BH}=A/(4 \tilde{G})=\frac{A}{4 (1+s/e^2)}$.
The tunneling rate is given by 
\begin{equation}\label{eq_tunneling_prob}
\Gamma=e^{-2 Im \hspace{2pt} \mathcal{A}}=e^{ \Delta S_{charged} }~,
\end{equation}
where 
\begin{equation}\label{eq_Stotal}
\Delta S_{charged}=\Delta S_{BH}+\Delta S_{\omega}~,
\end{equation}
is the total change in entropy of the scalar-hairy RN black hole due to the emission of massive charged particle.

For the emission of uncharged particles, $\Delta S_{\omega}=0$ and we get the tunneling rate as
\begin{equation}\label{eq_tunneling_prob_massless_uncharged}
\begin{split}
\Gamma&=e^{\Delta S_{BH}} \quad \mbox{with} \\ 
\Delta S_{BH}&=\frac{\pi}{1+s/e^2}\Bigg[\Bigg( (M_{ADM}-\omega ) \left(\frac{s}{e^2}+1\right)\\ 
 & \left.+\sqrt{(M_{ADM}-\omega )^2 \left(\frac{s}{e^2}+1\right)^2-e^2-s}\right)^2  \\
 &\left. - \left(M_{ADM}\left(\frac{s}{e^2}+1\right)+\sqrt{M_{ADM}^2\left(\frac{s}{e^2}+1\right)^2-e^2-s}  \right)^2\right].
\end{split}
\end{equation}
In case of vanishing scalar hair, $s=0$, we recover the tunneling rate for standard Reissner-Nordstr\"{o}m black hole \cite{parikh_PRL_2000, zhang_JHEP_2005}.

Expanding both the terms on the RHS of Eq.\ \eqref{eq_ImS_corr} to leading orders in $\omega$ and $q$,

\begin{equation}
\begin{split}
\Delta S_{BH}&=-\frac{2 \pi \left( M_{ADM} \left(\frac{s}{e^2}+1\right)+\sqrt{M_{ADM}^2 \left(\frac{s}{e^2}+1\right)^2-e^2-s} \right)^2}{\sqrt{M_{ADM}^2 \left(\frac{s}{e^2}+1\right)^2-e^2-s}}\\
& \times \Bigg[ \omega -\frac{e q}{M_{ADM} \left(\frac{s}{e^2}+1\right)+\sqrt{M_{ADM}^2 \left(\frac{s}{e^2}+1\right)^2-e^2-s}} \\
&-\frac{q s M_{ADM} }{e\left(e^2+s \right)} \Bigg]~,
\end{split}
\end{equation}
\begin{equation}
\begin{split}
\Delta S_{\omega}&= -2 \pi\frac{q s M_{ADM}}{e\left(e^2+s \right)\sqrt{M_{ADM}^2 \left(\frac{s}{e^2}+1\right)^2-e^2-s}} \\
& \times  \left( M_{ADM} \left(\frac{s}{e^2}+1\right)+\sqrt{M_{ADM}^2 \left(\frac{s}{e^2}+1\right)^2-e^2-s} \right)^2,
\end{split}
\end{equation}
we get the tunneling rate as
\begin{equation}\label{eq_thermal_prob}
\Gamma \sim e^{ -\beta\left(\omega-\omega_0\right) } ~,
\end{equation}
where $\beta=1/T_{BH}$ is the inverse of the black hole temperature, 
\begin{equation}
\beta=\frac{2 \pi \left( M_{ADM} \left(\frac{s}{e^2}+1\right)+\sqrt{M_{ADM}^2 \left(\frac{s}{e^2}+1\right)^2-e^2-s} \right)^2}{\sqrt{M_{ADM}^2 \left(\frac{s}{e^2}+1\right)^2-e^2-s}}
\end{equation}
and
\begin{equation}
\omega_0=\frac{e q}{ M_{ADM} \left(\frac{s}{e^2}+1\right)+\sqrt{M_{ADM} \left(\frac{s}{e^2}+1\right)^2-e^2-s}}~.
\end{equation}
As Eq.\ \eqref{eq_thermal_prob} includes only the leading order, Hawking radiation has an approximately thermal spectrum.

\begin{figure*}[!htbp]\centering
\includegraphics[width=0.95\linewidth]{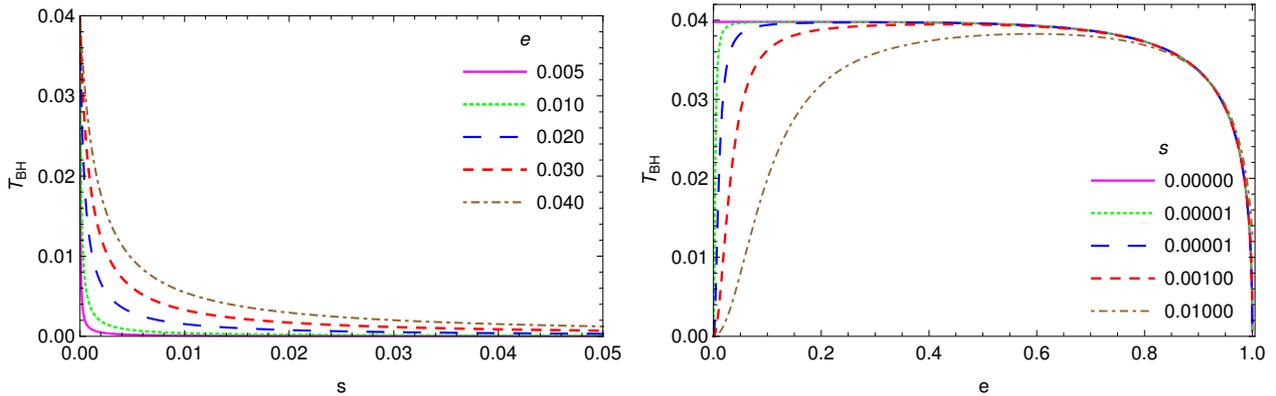}
\caption{{\bf (Left)} Plot of the Hawking temperature $T_{BH}$ of the scalar-hairy RN black hole with respect to the scalar charge $s$ for different values of the electric charge $e$  with $M_{ADM}=1$~. {\bf (Right)} Plot of the Hawking temperature $T_{BH}$ of the scalar-hairy RN black hole with respect to the electric charge $e$ for different values of the scalar charge $s$  with $M_{ADM}=1$~.}\label{fig:1}
\end{figure*}

Fig.\ref{fig:1}  shows the plot of the black hole temperature  with respect to the scalar and electric charges.

From Fig.\ref{fig:1} we observe that for a constant electric charge,  the black hole temperature $T_{BH}$  decreases very rapidly with $s$ but reaches a plateau quickly. For non-zero scalar charge, the black hole temperature rises with the electric charge, reaches a plateau and sharply falls to zero as the extremal value of the electric charge, $e_{ext}={\sqrt{\sqrt{M_{ADM}^2 \left(M_{ADM}^2+4 s\right)}+M_{ADM}^2}}/{\sqrt{2}}$~, is approached.  At sufficiently small values of $e$, $T_{BH}$ becomes vanishingly small.
\section{Charge-mass ratio bound from Mutual information of successively emitted charged Hawking quanta} \label{sec:charge-mass_ratio}
As already seen in Sec.\ref{sec:tunneling rate}, the Hawking radiation spectrum is not strictly thermal,  the emission in each step depends on the previous one. A quantity of importance in such a scenario is the ``Mutual Information", as defined in Refs.~\cite{zhang_PLB_2009, kim_PLB_2014},
\begin{equation}
\begin{split}
S_{MI}= & \Delta S_{charged}\left(M_{ADM},e;\omega_2,q_2|\omega_1,q_1\right)\\
&-\Delta S_{charged}\left(M_{ADM},e;\omega_2,q_2\right),
\end{split}
\end{equation}
where $\omega_1$, $\omega_2$and $q_1$, $q_2$ are the mass and charge of two consecutive emissions, $\Delta S_{charged}\left(M_{ADM},e;\omega_2,q_2|\omega_1,q_1\right)$ is the change in entropy of the black hole due to the emission ofrticle of mass $\omega_2$ and charge $q_2$ considering a previous emission of mass $\omega_1$ and charge $q_1$.
$S_{MI}$ gives a measure of the correlation between two consecutive emission. The conservation of energy in the tunneling method automatically ensures the conservation of information\cite{zhang_PLB_2009}. The non-negativity of mutual information during emission of two successive Hawking quanta gives rise to bounds on the charge to mass ratio of the emitted particles.

Far away from extremality, in the non superradiant regime, in the limit of large  black hole mass  and charge, $M_{ADM}\gg \left\{\omega_1, \omega_2\right\}$; $e\gg \left\{q1,q2\right\}$ and $M_{ADM}\gg e,s$, the non-negativity condition yields,
\begin{equation}
\begin{split}
 \frac{q_i}{\omega_i}\leq  \frac{1}{e^3}\Big[ &\sqrt{2}\sqrt{2s^2\left( M_{ADM}-\omega_i \right)^2+e^4 \left( s+e^2  \right)}\\
&-2s \left( M_{ADM}-\omega_i \right) \Big],
\end{split}
\end{equation}
which can be written as
\begin{equation}\label{eq_q/w}
\frac{q_i}{\omega_i}\leq-\frac{2 s M_{ADM}}{e^3}+\sqrt{ \left( \frac{2 s M_{ADM}}{e^3} \right)^2+2\left(1+s/e^2\right) }~,
\end{equation}
where, for convenience, we have chosen $\omega_1=\omega_2=\omega_i$ and $q_1=q_2=q_i$ .
In the absence of the scalar hair ($s=0$),  Eq.\ \eqref{eq_q/w} reduces to $q_i/\omega_i\leq \sqrt{2}$~, which matches exactly with the charge-mass ratio for the RN black hole, obtained in Ref.~\cite{kim_PLB_2014}.
In order to obtain Eq.\ \eqref{eq_q/w}, we expanded the integrand in Eq.\ \eqref{eq_Somega} to leading orders in $\omega$ and $q'$  as the integration is otherwise difficult to perform.
\begin{figure*}[!htp] \centering
\includegraphics[width=0.98\linewidth]{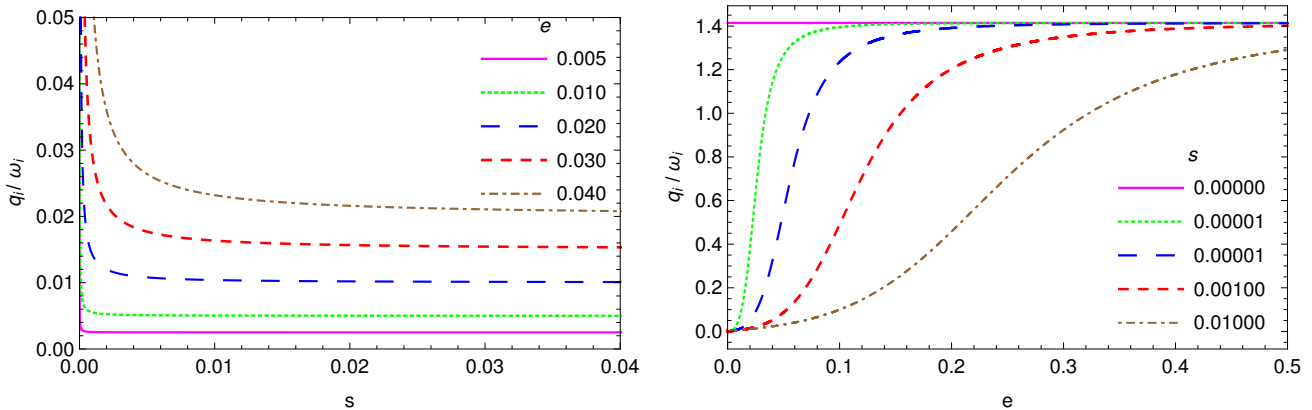}
\caption{{\bf (Left)} Plot of the maximum charge-mass ratio with $s$ for different values of the electric charge $e$ with $M_{ADM}=1$~. {\bf (Right)} Plot of the maximum charge-mass ratio with $e$ for different values of the scalar charge $s$ with $M_{ADM}=1$~.}\label{fig:2}
\end{figure*}

Fig.\ref{fig:2} shows the variation of the upper bound on the charge-mass ratio with the scalar and electric charges of the black hole. We observe that at a fixed electric charge of the black hole, for smaller values of the scalar charge, the upper bound on the charge-mass ratio decreases with $s$~; however, the rate of fall decreases with the increase in the scalar charge. On the other hand, for non-zero scalar charge, the maximum possible charge-mass ratio increases with the electric charge of the black hole  and approaches that for the standard RN black hole.
\section{Summary and Discussion}\label{sec:summary}
In this paper, we studied the problem of Hawking emission of massive charged particles from a spherically symmetric charged black hole endowed with a scalar hair using the semi-classical tunneling formalism. The black hole is very similar to the standard Reissner-Nordstr\"{o}m black hole  but with an additional scalar hair (See Refs.\citep{astorino_PRD_2013, paper1_EPJC_2018, paper2_arxiv_2019}). The presence of the scalar hair gives rise to an effective Newtonian constant, $\tilde{G}=G\left( 1+s/e^2 \right)$ and an ADM mass, different from the mass parameter $M$ (see Eq.\ \eqref{eq_MADM}).
It is interesting to note that in the limit of vanishing electric charge,  the ADM mass of the black hole goes to zero which is due to the fact that the effective Newtonian constant $\tilde{G}$ blows up as $e\rightarrow0$. This suggests that the scalar-hairy RN black hole cannot radiate away its electric charge and settle down to a scalar-hairy (electrically) uncharged distribution. This is consistent with the following thermodynamic consideration. We note that for any positive value of the scalar charge, the black hole temperature becomes vanishingly small at sufficiently small values of the black hole electric charge (see Fig.\ref{fig:1}) and thus according to the Third Law of black hole thermodynamics\cite{israel_PRL_1986}, it is impossible for the scalar-hairy RN black hole to radiate away its electric charge in any finite number of steps.
  
The total change in entropy of the scalar-hairy RN black hole due to the emission of the massive charged particle contains a $\omega$-dependent contribution due to the scalar charge (see Eqs. \eqref{eq_Somega},~\eqref{eq_Sbh} and \eqref{eq_Stotal}). The emission rate, (see Eq.\ \eqref{eq_tunneling_prob}) matches smoothly with that for the standard Reissner-Nordstr\"{o}m black hole (see Ref.~\cite{zhang_JHEP_2005}) in the limit of vanishing scalar charge $s$. The emission rate for uncharged massless particles obtained by putting $q=0$ in Eq.\ \eqref{eq_tunneling_prob} (see Eq.\ \eqref{eq_tunneling_prob_massless_uncharged}) is also of the same functional form as that derived using the geodesic equation. We also note that the Hawking emission spectrum deviates from pure thermality. This is consistent with the generic results found in the investigations~\cite{parikh_PRL_2000, parikh_IJMPD_2004, parikh_GRG_2004, parikh_MARCELGROSSMANN10, zhang_JHEP_2005, jiang_PLB_2006, liu_IJTP_2014, jiang_PRD_2006, jiang_IJTP_2006, jiang_PLB_2008, sarkar_PLB_2008, wu_JHEP_2006, li_PLB_2006, chen_PLB_2008, chen_CQG_2008, jiang_PRD_2008, jiang_JCAP_2013, zhang_PLB_2005} and the recent findings that the quantum process involved in Hawking radiation is unitary, meaning a pure state to pure state transition\cite{chen_PLB_2009,singleton_JHEP_2010,sakalli_ASS_2012,saini_PRL_2015, arpit_EPJC_2019}.

It is important to note that using the modified geodesic equation instead of the Eq.\ \eqref{eq_vp_vg} as suggested by Pu and Han\cite{pu_IJTP_2017} also yields the same expression of the tunneling rate.

We also studied the mutual information stored in consecutive Hawking emission. Demanding non-negativity of the mutual information, we observed that the maximum allowed charge-mass ratio of the emitted particles decreases with the scalar charge of the black hole. 

It will be interesting to study the effect of Lorentz symmetry breaking on quantum tunneling in other black hole solutions such as those in bumblebee gravity. A few investigations have already been done in this direction (see Ref.~\cite{casana_PRD_2018, kanzi_NPB_2019,ovgun_PRD_2019}).

\begin{acknowledgements}
The author is indebted to Professor Narayan Banerjee for valuable discussions and comments.
\end{acknowledgements}
\bibliographystyle{apsrev4-1}
\bibliography{ref3}
\end{document}